\documentclass[twocolumn,prl,aps,superscriptaddress,showpacs]{revtex4}

\usepackage{bm}\usepackage{graphicx}

\begin{document}

\title{Low energy magnetic excitation spectrum of the unconventional ferromagnet CeRh$_{3}$B$_{2}$}

\author{S. Raymond}
\affiliation{SPSMS, UMR-E 9001, CEA-INAC/ UJF-Grenoble 1, 38054 Grenoble, France} 
\author{J. Panarin}
\affiliation{SPSMS, UMR-E 9001, CEA-INAC/ UJF-Grenoble 1, 38054 Grenoble, France}
\author{F. Givord}
\affiliation{SPSMS, UMR-E 9001, CEA-INAC/ UJF-Grenoble 1, 38054 Grenoble, France}
\author{A.P. Murani} 
\affiliation{Institut Laue Langevin, 38042 Grenoble Cedex, France}
\author{J.X. Boucherle}
\affiliation{Institut N\'eel, CNRS, 38042 Grenoble Cedex, France}
\author{P. Lejay}
\affiliation{Institut N\'eel, CNRS, 38042 Grenoble Cedex, France}
             
\date{\today}

\begin{abstract}
The magnetic excitation spectrum of the unconventional ferromagnet CeRh$_{3}$B$_{2}$ was measured by inelastic neutron scattering on single crystal sample in the magnetically ordered and paramagnetic phases. The spin-wave excitation spectrum evidences high exchange interaction along the $c$-axis about two orders of magnitude higher than the ones in the basal plane of the hexagonal structure. Both strong out of plane and small in plane anisotropies are found. This latter point confirms that considering the  $J$=5/2 multiplet alone is not adequate for describing the ground state of CeRh$_{3}$B$_{2}$. Quasielastic scattering measured above $T_{Curie}$ is also strongly anisotropic between the basal plane and the $c$-axis and suggests localized magnetism.
\end{abstract}

\maketitle

\section{Introduction}
 CeRh$_{3}$B$_{2}$ is a Ce-based intermetallic compound with quite unique properties. It is ferromagnetic while most of the other Ce-based compounds are antiferromagnetic. The easy axis is in the basal plane of the hexagonal structure and the saturation magnetization 0.4 $\mu_{B}$ is strongly reduced compared to free cerium ion value (2.14 $\mu_{B}$). More striking is its huge Curie temperature, $T_{Curie}$=115 K, that is two orders of magnitude higher that what would be expected from applying the de Gennes scaling to GdRh$_{3}$B$_{2}$ ($T_{Curie}$=90 K) \cite{Dhar}. Different theoretical models from itinerant to localized magnetism have been proposed to explain the peculiarities of this compound and they are reviewed in the recent calculations \cite{Yamauchi,Kono}. 
The key ingredient is certainly the very short Ce-Ce distance (3.08 $\AA$) along the $c$-axis that leads to the proposition of strong interatomic Ce-4$f$-Ce-5$d$ hybridization or even direct Ce-4$f$-Ce-4$f$ interaction \cite{Kasu}. While initially itinerant magnetism was favoured, many experimental results point toward localized 4$f$ electrons with strong hybridization as inferred from photoemission spectroscopy \cite{Fujimori}, NMR \cite{NMR}
dHvA measurements \cite{DHVA}. Polarized neutron diffraction and crystal field spectroscopy give an evidence of a mixing of $J$=5/2 and $J$=7/2 multiplets in the ground state wave-function \cite{Givord1,Givord3}.
Compton scattering experiments \cite{Yaouanc,Sakurai} show that the orbital to spin moment  ratio is less than the expected value for Ce$^{3+}$ ion. Hence the spin magnetism is enhanced in CeRh$_{3}$B$_{2}$ with respect to the orbital part and this could explain partly the high Curie temperature since exchange interactions couple spins. These overall features together with the strong recent interest for other $f$ electron ferromagnetic systems, namely the new uranium based ferromagnetic superconductors \cite{ferrosupra}, motivated us to investigate the spin dynamics of CeRh$_{3}$B$_{2}$. It is worthwhile to note that no trace of superconductivity was detected when ferromagnetism is suppressed at 7 GPa \cite{Cornelius} in CeRh$_{3}$B$_{2}$ while superconductivity is known to exist in Ce(Rh$_{1-x}$Ru$_{x}$)$_{3}$B$_{2}$ for 0.62 $<$ $x$ $<$ 1 and ferromagnetism disappears for $x$ $>$ 0.16 \cite{Allen}.

\section{Experimental details}
Experiments were carried out on the three axis spectrometers IN12, IN22, IN20 and IN8 at Institut Laue Langevin, Grenoble. The configuration of each spectrometer is given in Table 1. For all configurations, a fixed final energy was used together with a focusing analyzer without collimation. Except for IN8 where measurements were performed around (0, 0, 2), most measurements were carried out around the very weak nuclear reflection (0, 0, 1) in order to limit the contamination by acoustic phonons. The single crystal was grown by the Czochralski method by using isotopically enriched $^{11}$B (90 \%) in order to reduce neutron absorption. The sample for neutron scattering consists of two co-aligned platelets with [0, 0, 1] normal.  Their thickness is 2 mm in order to reduce the effect of Rh neutron absorption. They are of typical length 25 mm and width 5 mm.
\begin{table}
\caption{Instrument configurations. The given collimation is the one after the monochromator. Fixed final wave-vector was used with a filter on the scattered beam. PG is pyrolitic graphite}
\begin{tabular}{cccccc}
\hline
  & Monoch. & Collim. & Anal. & Wave-vector & filter\\
 \hline
 IN12 & PG & 60' & PG & 1.5 & Be \\
 IN22 & PG & 60' & PG & 1.97 , 2.662 & PG\\
 IN20& Si & - & PG & 2.662&-\\
 IN8& Si & - & PG & 4.1 & PG\\
 IN8& Cu & - & PG & 4.1 & PG\\
\end{tabular}
\label{table}
\end{table}

\begin{figure}
\includegraphics[width=8cm]{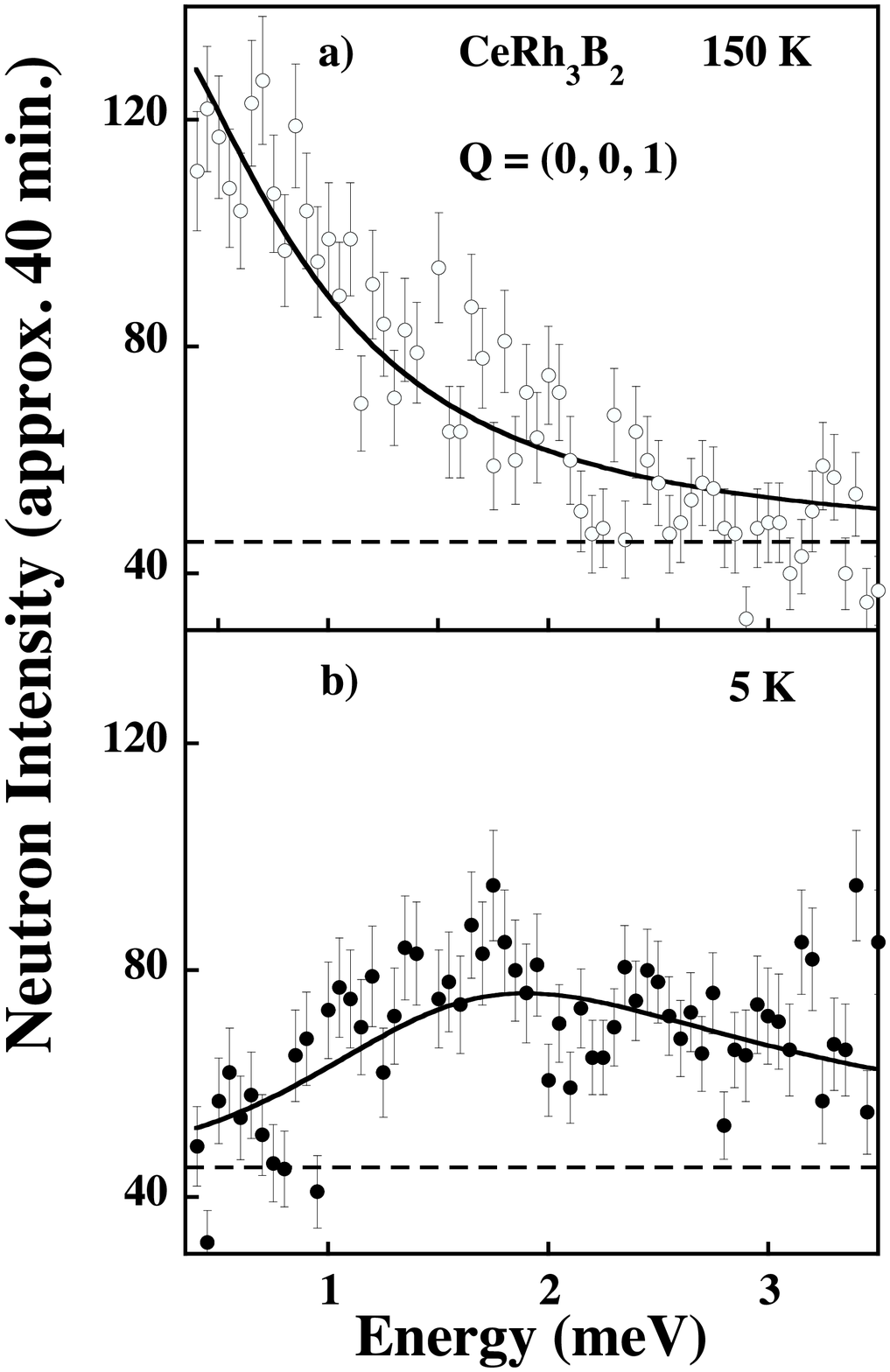}
\caption{Magnetic excitation spectrum of CeRh$_{3}$B$_{2}$ at $\bf{Q}$=(0, 0, 1) for a) T = 150 K and b) 5 K. Solid lines are fit to the data with a quasielastic Lorentzian at 150 K and an inelastic Lorentzian at 5 K (See Appendix). The dashed line indicates the background.}
\end{figure}

\section{Spin-waves}
Figure 1 shows magnetic excitation spectrum measured on IN12 at $\bf{Q}$=(1, 0, 0) for T = 5 and 150 K. They were fit with inelastic Lorentzian below $T_{Curie}$ and quasielastic Lorentzian above $T_{Curie}$ (See Appendix). These spectra give evidence of a gap in the excitation spectrum of about 2 meV for $\bf{q}$=0 in the ferromagnetic phase (In this paper, $\bf{Q}$=$\tau$+$\textbf{q}$ where $\tau$ is a zone center and $\bf{Q}$=(Q$_{H}$, Q$_{K}$, Q$_{L}$) and $\bf{q}$=(h, k, l)). The dispersion along the [1, 0, 0] direction was investigated by performing several constant $\bf{Q}$ scans. Figure 2 shows a representative magnetic excitation spectrum measured on IN20 at $\bf{Q}$=(0.1, 0, 1) for 5 and 150 K. In this work, the background is consistently taken for each configuration as the flat part of the constant energy and constant $\bf{Q}$ scans. The dispersion along the [0, 0, 1] direction was measured by performing constant energy scans since the dispersion is much steeper in this direction than along [1, 0, 0]. Representative spectra are shown in Figure 3. Figure 3a shows measurements performed on IN20 for an energy transfer of 9 meV subtracted from the background and corrected for the Bose factor. In the ferromagnetic phase, the asymmetry between the two peaks located on both sides of (0, 0, 1) corresponds to the defocusing and focusing conditions for the measurements. The data in the paramagnetic phase exhibits weak $\bf{Q}$-dependence due to magnetic correlations (see next part). Figure 3b shows a subtraction of the data measured at 5 K and 150 K for an energy transfer of 40 meV on IN8. For this measurement the background is not well determined due to optic phonon and multi-phonon contamination near (0, 0, 2) so that the same procedure as the one shown in Fig.3a is not applied. However the peak position is still well-defined. The resulting dispersion is shown in Figure 4. There are three salient features : (i) the small gap at $\bf{q}$=0, (ii) the linear dispersion at small $\bf{q}$ and the very high anisotropy of the spin-wave energy between the two directions. Without further analysis these facts already allow to draw some conclusions on the spin dynamics of CeRh$_{3}$B$_{2}$. Indeed for a ferromagnetic planar system, the spin-wave spectrum is linear and gapless. The observed gap is thus a signature of the in-plane anisotropy. Therefore, the spin wave is analyzed by using the following dispersion relation \cite{Nira} :
\begin{equation}
\omega^2_{q}=\left[\Delta_{1}+2SI(0)-2SI(\bf{q})\right]\left[\Delta_{2}+2SI(0)-2SI(\bf{q})\right]
\end{equation}
$\omega_{q}$ in the energy of the mode, $\Delta_{1}$ is the axial anisotropy and $\Delta_{2}$ is the in-plane anisotropy. $I(\bf{q})$ is the Fourier transform of the exchange integral $I_{i,j}$ between the magnetic sites $i$ and  $j$ with the Hamiltonian $\sum_{i,j}-I_{i,j}\textbf{S}_{i}\textbf{S}_{j}$. $\textbf{S}_{i}$ is the spin operator at site $i$ and $S$ is the value of the spin. Table II shows the different exchange integrals and the corresponding distances between Ce atoms and the number of neighbors $z$.
Given the shape of the dispersion, the relevant exchange parameters that we have deduced are : $I_{c}$ the nearest neighbor interaction along the $c$-axis, $I_{1}$ the nearest neighbor interaction in the basal plane and $I_{2}$, the next nearest neighbor interaction in the plane. Note that because measurement are performed only along [1, 0, 0] and [0, 0, 1], the exchange $I_{diag}$ along [1, 0, 1] is included in $I_{c}$ and $I_{1}$. If $I_{diag}$ is strong (which is not really expected because the corresponding Ce-Ce distance is high), $I_{c}$ and $I_{1}$ are effective parameters. We cannot also determine separately  the exchange parameters and the energy gaps because they occurs as products in Eq.(1).
The obtained parameters from a fit of the dispersion shown in Fig.4 are given in Table II. $I_{1}$ is not very well determined because part of $I_{2}$ intervenes in the same way as $I_{1}$ in the fit. The salient feature is the large difference between $I_{c}$ and $I_{1}$ ($I_{1}$), the former being  two orders of magnitude larger than the latter : $I_{c}/I_{1} \approx$ 300 and $I_{c}/I_{2} \approx$ 100. The dispersion along the [0, 0, 1] direction is characteristic of the regime $\Delta_{1}$ $>$ 2$SI_{c}$ and good fits are indeed only obtained for $\Delta_{1}$ $>$ 100 meV that leads also to the constraint 2$SI_{c}$ $<$ 20 meV. We will tentatively give individual estimate for $\Delta_{1}$ and $I_{c}$ in the discussion part with extra hypotheses beyond the present fit. The gap at $\textbf{q}$=0 is the geometric mean of the axial and in-plane anisotropies: $\omega_{q=0}$=$\sqrt{\Delta_{1}\Delta_{2}}$ $\approx$ 2 meV.
\begin{table}[b!]
\caption{Exchange integrals and anisotropies}
\begin{tabular}{cccccc}
\hline
- & distance & direction & z  & Fit Results  &  \\
 \hline
 $I_{1}$&5.48&[1,0,0]&6&$\Delta_{1}2SI_{1}$ & 8(8) meV$^{2}$\\
 $I_{2}$&9.49&[1,1,0]&6&$\Delta_{1}2SI_{2}$ & 19 (7) meV$^{2}$\\
 $I_{c}$&3.04&[0,0,1]&2&$\Delta_{1}2SI_{c}$ & 2520 (40) meV$^{2}$\\
 $I_{c'}$&6.08&[0,0,1]&2&-\\
 $I_{diag}$&6.26&[1,0,1]&6&-\\
 \hline
 $$&$$&$$&$$&$\Delta_{1}\Delta_{2}$& 5(2) meV$^{2}$\\
 \hline
\end{tabular}
\label{table}
\end{table}

\begin{figure}
\includegraphics[width=8cm]{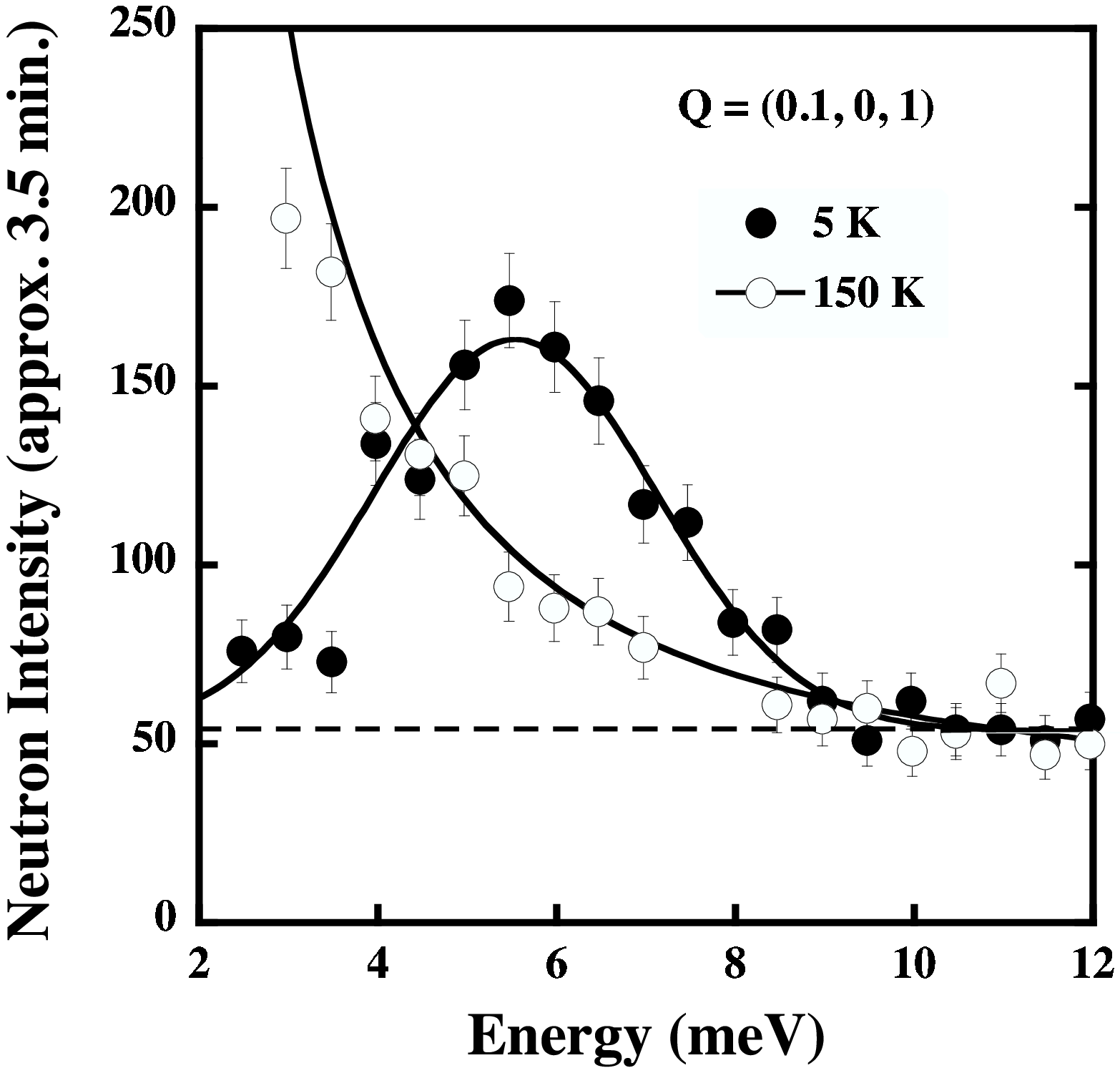}
\caption{Magnetic excitation spectrum of CeRh$_{3}$B$_{2}$ at $\bf{Q}$=(0.1, 0, 1) for T = 5 and 150 K. Solid lines are fit to the data with a quasielastic Lorentzian at 150 K and a Gaussian at 5 K (See Appendix). The dashed line indicates the background.}
\end{figure}

\begin{figure}
\includegraphics[width=8cm]{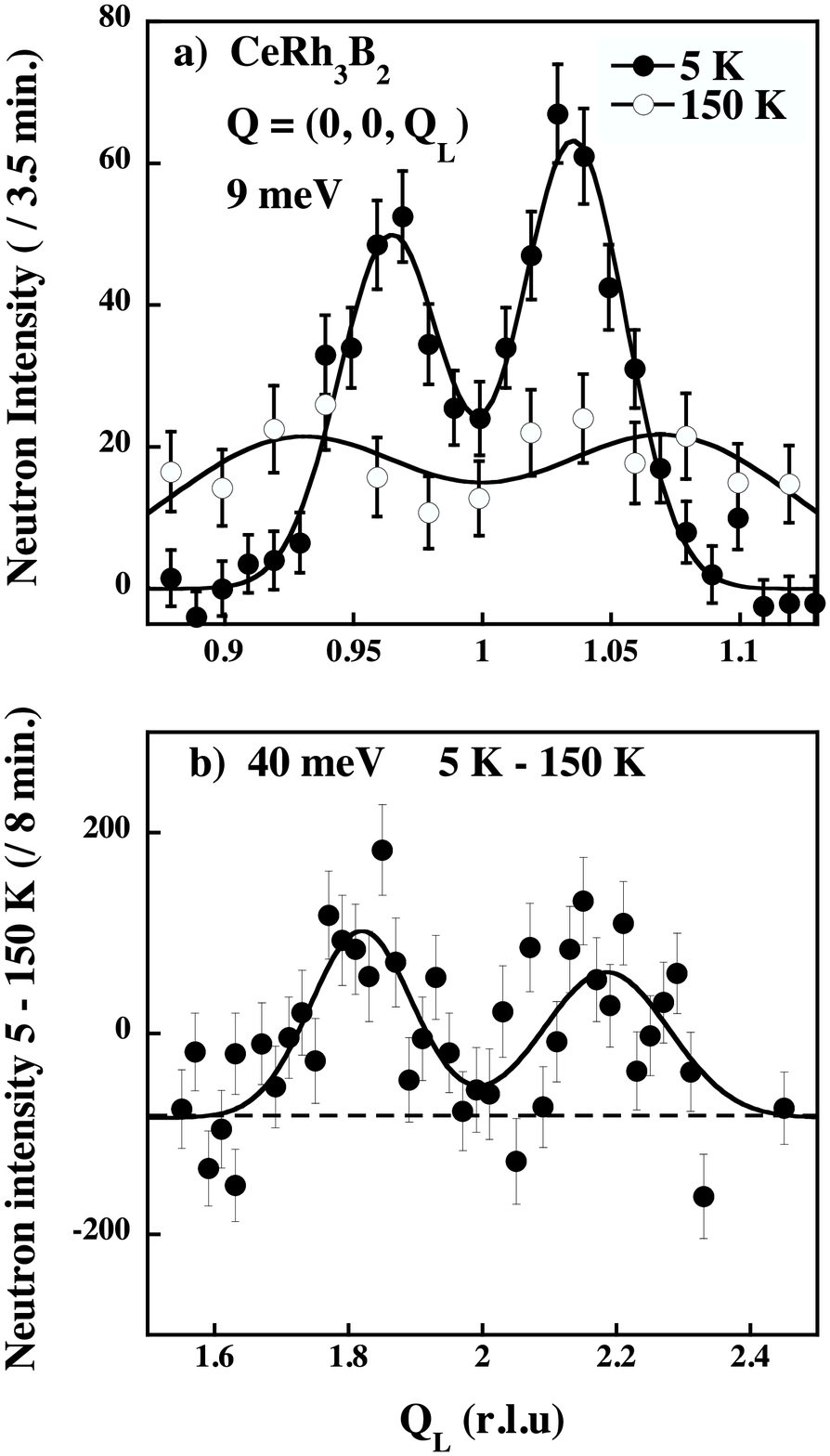}
\caption{a) Constant energy scans performed along [0, 0, 1] at 9 meV at 5 K and 150 K with the background subtracted and corrected by the Bose factor. b) Subtraction of the raw data measured at 5 and 150 K by constant energy scan along [0, 0, 1] with energy transfer of 40 meV. Lines are fits with Gaussians (see Appendix).}
\end{figure}

\begin{figure}
\includegraphics[width=8cm]{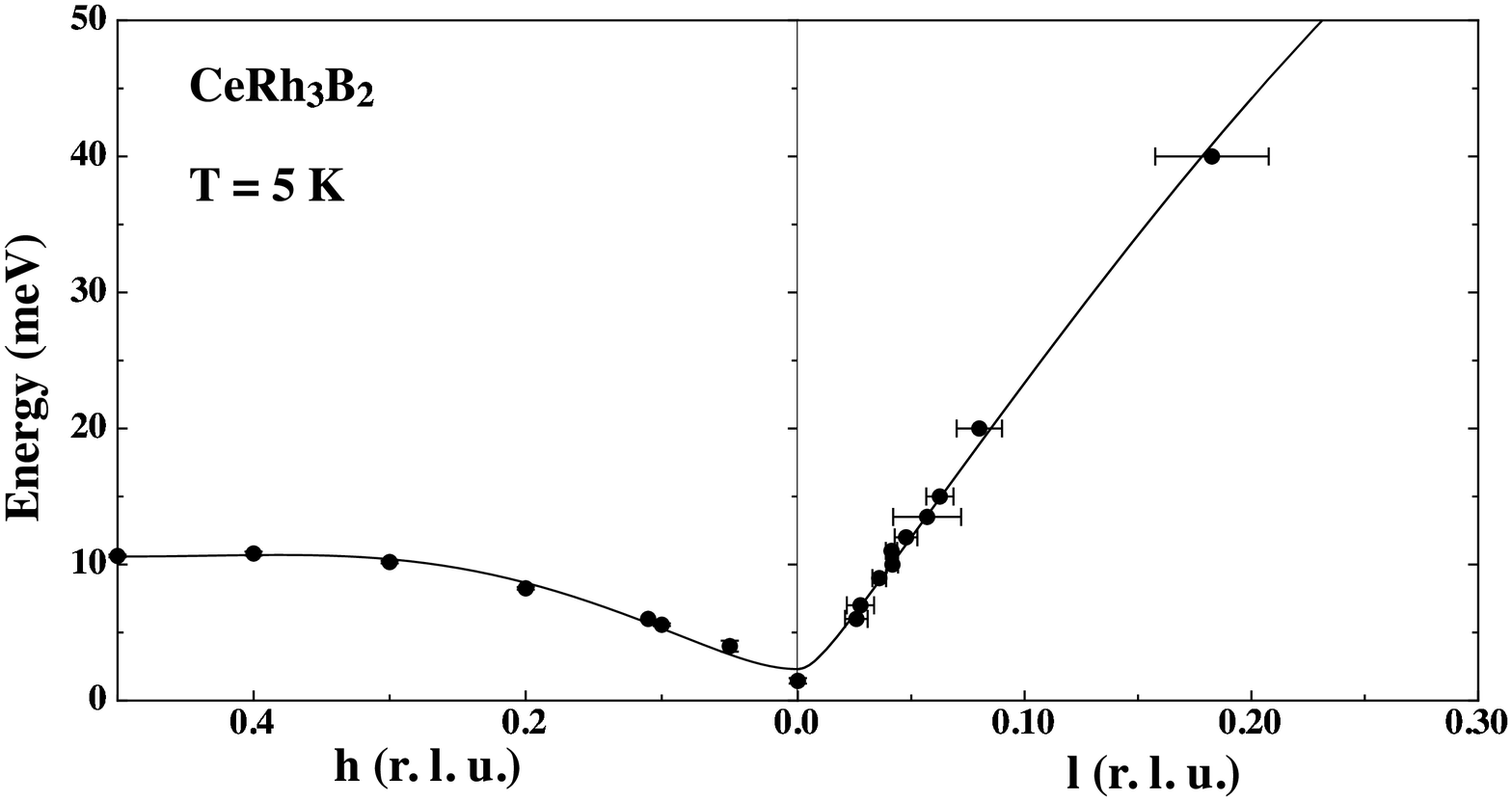}
\caption{Spin-wave dispersion in the [1, 0, 0] and [0, 0, 1] directions at 5 K. Lines are fit as explained in the text.}
\end{figure}

\section{Paramagnetic scattering}

\begin{figure}
\includegraphics[width=8cm]{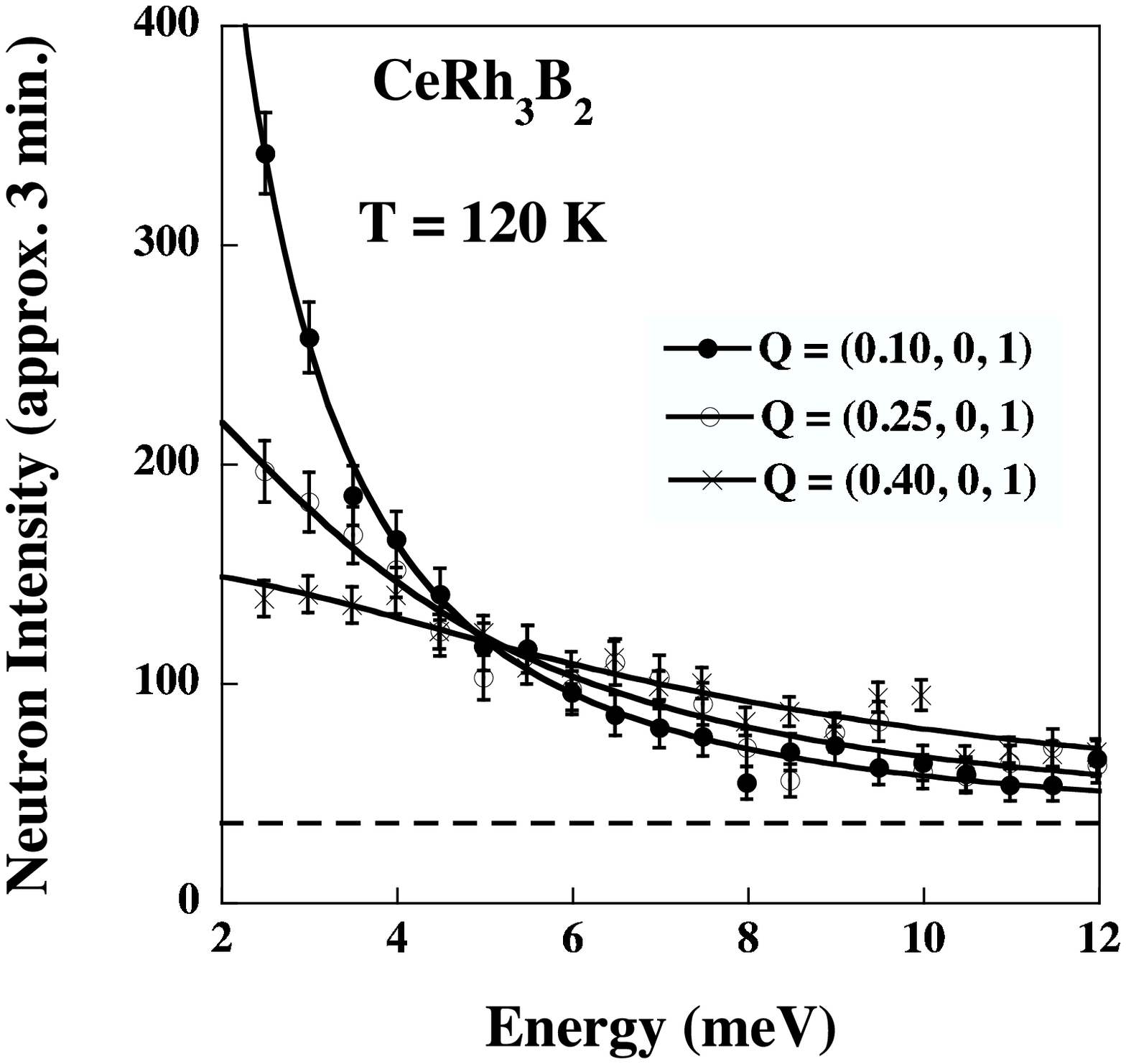}
\caption{Energy spectra measured at 120 K. Solid lines are fit to the data with a quasielastic Lorentzian (See Appendix). The dashed line indicates the background.}
\end{figure}
\begin{figure}
\includegraphics[width=8cm]{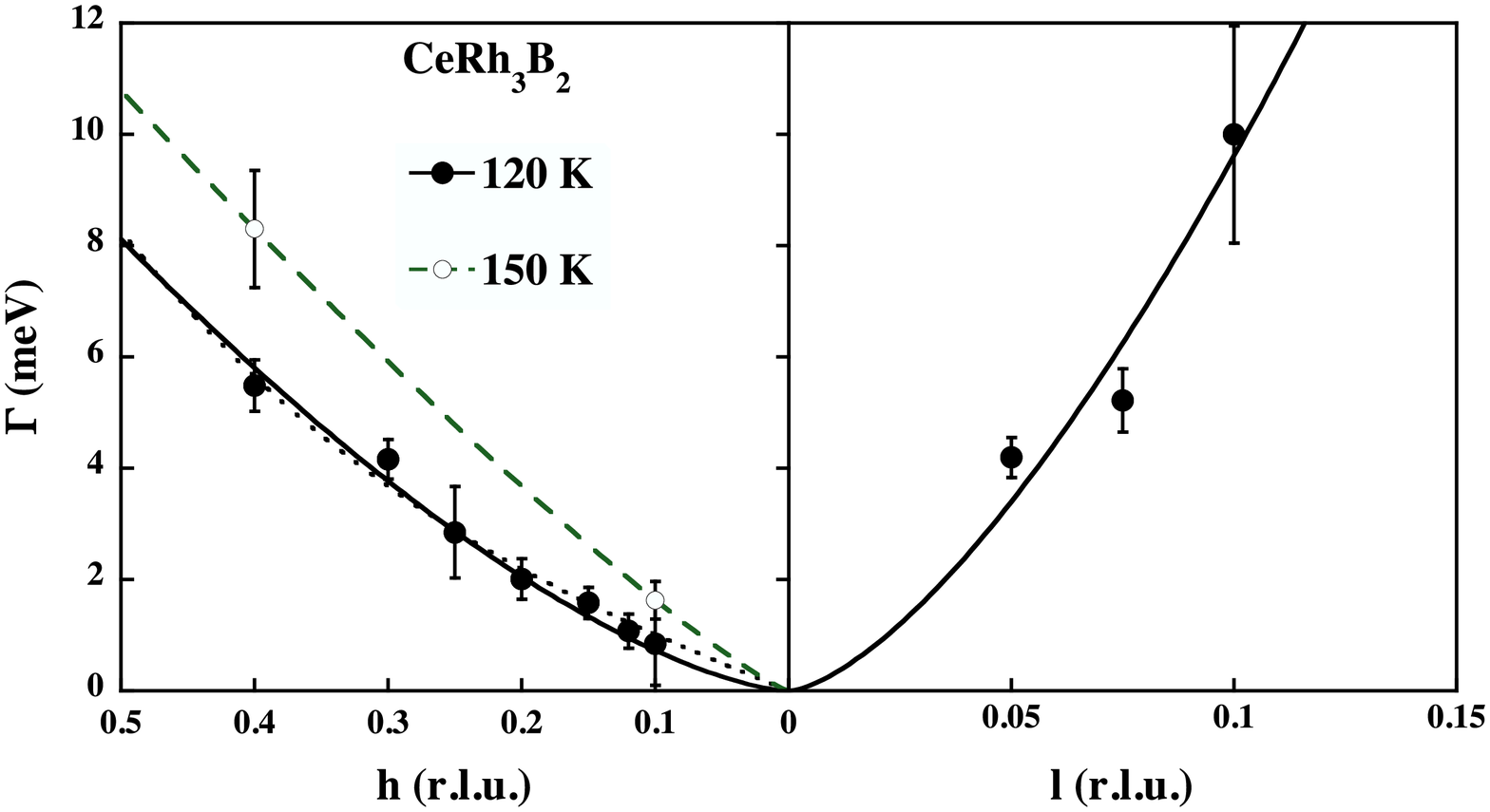}
\caption{Relaxation rate in the [1, 0, 0] and [0, 0, 1] directions at 120 K and 150 K. Lines are fit as explained in the text ; The solid line corresponds to the local model and the dotted line corresponds to the itinerant model. The dashed line is a guide for the eyes.}
\end{figure}

\begin{figure}
\includegraphics[width=8cm]{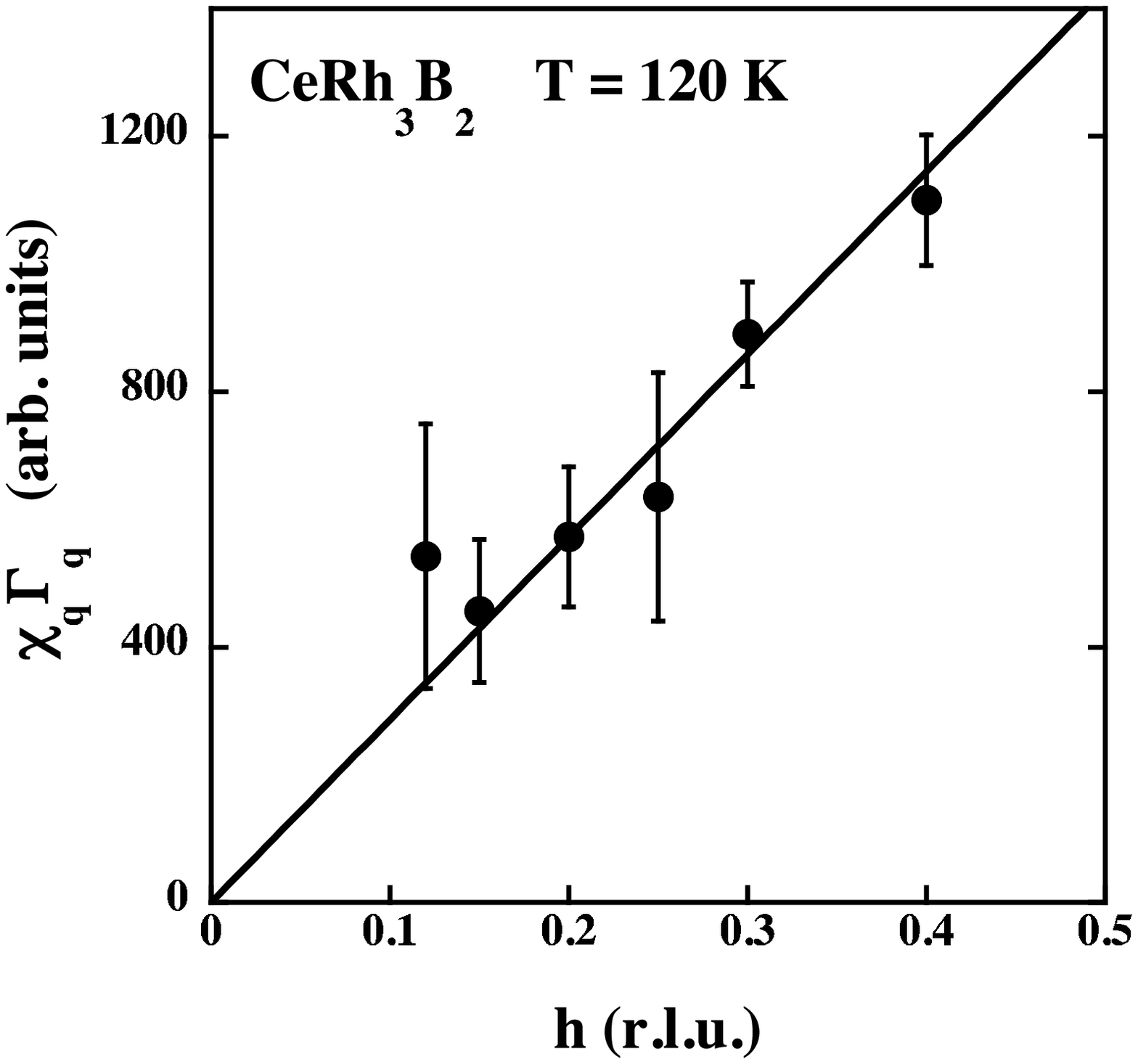}
\caption{The product $\chi_{q}\Gamma_{q}$ at 120 K for $\textbf{q}$ along the [1, 0, 0] direction. The line is a linear fit going through the origin.}
\end{figure}

Preliminary data concerning paramagnetic scattering obtained on powder sample were shown in Ref.\cite{Givord3}. A quasielastic signal was observed with a powdered average linewidth of 2 meV at 150 K and 5 meV at 300 K. 
Figure 5 shows representative spectra taken in the present single crystal study at 120 K just above $T_{Curie}$  for three wave-vectors, which establishes the $\bf{Q}$ dependence of the signal. The data are analyzed with a quasielastic Lorentzian lineshape convoluted with one dimensional resolution in the $\omega$ direction. This analysis gives two parameters, the $\textbf{q}$-dependent susceptibility $\chi_{q}$ and the relaxation rate $\Gamma_{q}$. In this work, the probed fluctuations are in-plane fluctuations since the wave-vector has predominant component along the $c$-axis (being (h, 0, 1) or (0, 0 1+l)) and neutron probes fluctuations perpendicular to $\textbf{Q}$. The obtained relaxation rate is shown in Figure 6 at 120 K and for [1, 0, 0] and [0, 0, 1] and for a few points at 150 K along [1, 0, 0]. The obtained values are qualitatively consistent with the powder averaged ones since the basal plane has more spectral weight than the $c$-axis direction. The extent in $\omega$ space of these excitations is smaller than that of the spin wave for the [0, 0, 1] direction and quite similar for the [1, 0, 0] direction. 
It seems that the relaxation rate extrapolates to zero for $\textbf{q}$=0. However this is not expected since the in-plane magnetization is not a conserved quantity in a planar magnet \cite{Thoma}.
Hence a small finite value $\Gamma_{q}(q=0)$ may exist but our data do not allow us to extract it.
We fit the data obtained along the [1, 0, 0] direction at 120 K with either $\Gamma$=$Dh^z$ or $\Gamma=ah(1+(h/\kappa)^2)$ that corresponds respectively to localized or itinerant model of ferromagnetism.
Along [1, 0, 0], we obtained $D$= 19(3) and $z$= 1.34(14) (solid line in Fig.6) and respectively $a$=9.9(9) meV.(r.l.u.)$^{-1}$ and $\kappa$=0.6(1) r.l.u. (dotted line in Fig.6). It is worthwhile to note that $z$ is close to 3/2, the dynamical exponent expected for an X-Y magnet \cite{Thoma}. The fit along the [0, 0, 1] direction is not exploited due to the limited number of data points.
Usually the ratio $D/T_{Curie}$ can give some information on the nature of the magnetism i.e. localized versus itinerant. It is expected that $<\Gamma_{q}>/T_{Curie}$ $>>$ 1 for itinerant magnetism and that $<\Gamma_{q}>/T_{Curie}$ $\approx$ 1 for localized magnetism \cite{Moriya}. This classification was applied for isotropic cubic magnets \cite{cubic} or more recently in orthorhombic UGe$_{2}$ \cite{Raymond}. It is highly questionable to apply it here since the value of $D$ is quite different between the plane and the $c$-axis (See Fig. 6). We could nevertheless notice that the powder averaged value of the relaxation rate \cite{Givord3} leads to $<\Gamma_{q}>/T_{Curie}$ $<$ 1 that is in favor of localized magnetism. This conclusion is also sustained by the value of $\kappa$ obtained by the itinerant model fit of $\Gamma_{q}$ that is quite large for a temperature just above the Curie temperature and this analysis seems therefore unphysical. Hence our limited set of data obtained in the paramagnetic phase would suggest localized magnetism.
Finally as compared to antiferromagnetic heavy fermion systems where $\chi_{q}\Gamma_{q}$ is constant, it is worthwhile to point out that here $\chi_{q}\Gamma_{q}$ is almost linear in $\bf{q}$ for the [1, 0, 0] direction at 120 K for which data were extensively taken (See Fig. 7). This emphases the peculiarity of ferromagnetic fluctuations and the almost conserved magnetization. In the ferromagnetic compound UGe$_{2}$, it was found that $\chi_{q}\Gamma_{q}$ is linear with a large finite intercept at $q$=0 \cite{Huxley}. Constant energy scans performed for an energy transfer of 3 meV are shown in Figure 8 for 120 and 200 K for [1, 0 ,0] and [0, 0, 1] directions. The peak width is very anisotropic between the basal plane where peaks from adjacent Brillouin zones overlap and the $c$-axis where peaks are narrow.
The constant energy scans can be fit by Lorentzian or Gaussian lineshape and both fits provide almost the same width $\kappa^{*}$. This is not the true inverse correlation length $\kappa$ since the measurement is performed for a finite energy of 3 meV while $\kappa$ is obtained by energy integrated imaginary part of the dynamical spin susceptibility. The gaussian fit gives $\kappa^{*}_{[1,0,0]}$=0.33(1) r.l.u. , $\kappa^{*}_{[0,0,1]}$=0.065(4) at 120 K and $\kappa^{*}_{[1,0,0]}$=0.38(5) , $\kappa^{*}_{[0,0,1]}$=0.100(8) at 200 K. The temperature dependence of this characteristic length is more important along the $c$-axis. This could explain the above mentioned fact concerning the extent in $\textbf{q}-\omega$ space of the fluctuations. The change of this extension along [0, 0, 1] between the paramagnetic and ordered phase could be due to a stronger temperature dependence of the different parameters in this direction. Exploiting further the paramagnetic scattering would need a detailed survey of the spin dynamics in the ($\bf{Q}$,$\omega$) space above $T_{Curie}$.

\begin{figure}
\includegraphics[width=8cm]{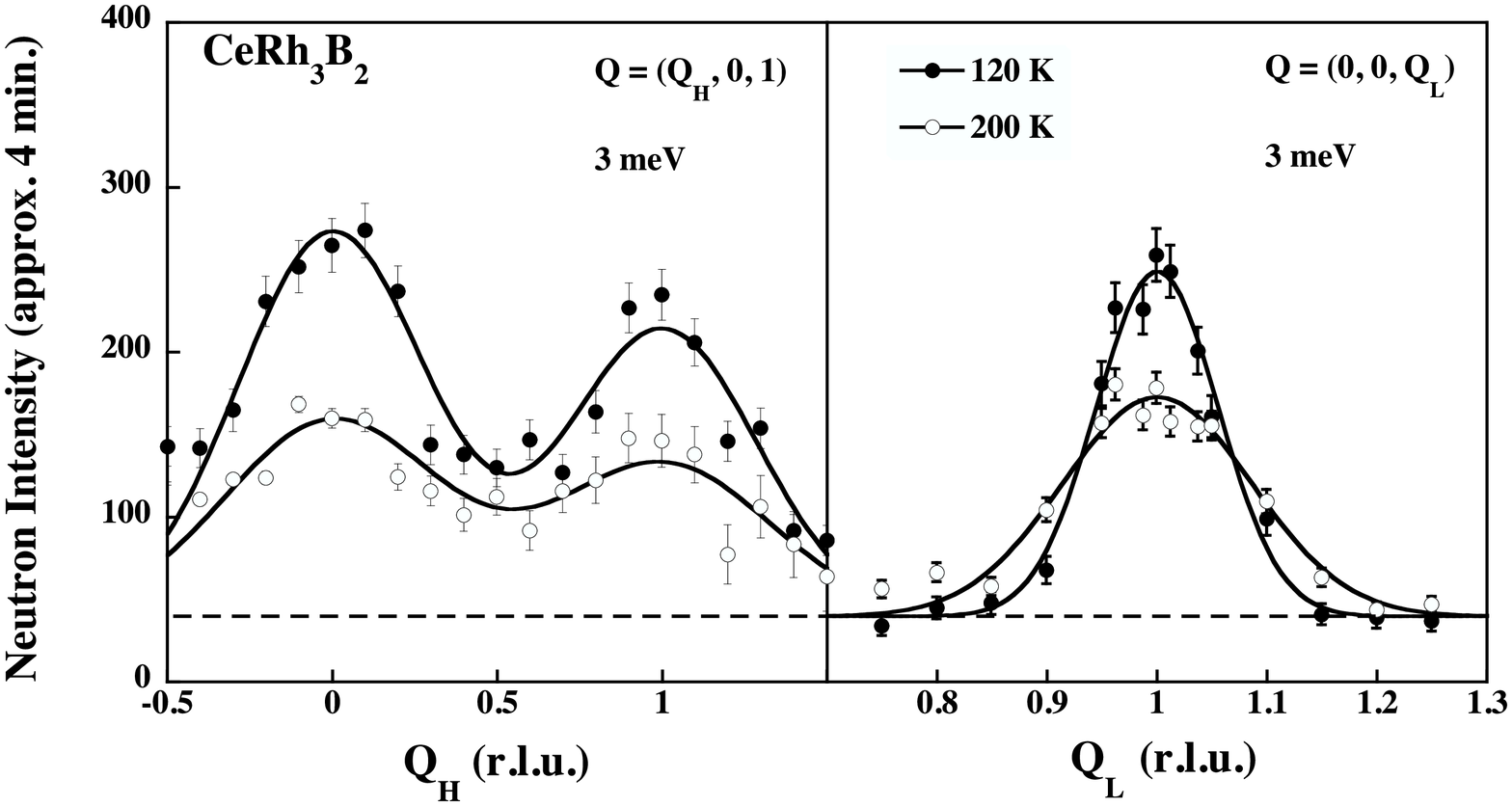}
\caption{Constant energy scans performed for an energy transfer of 3 meV at 120 and 200 K along [1, 0, 0] and [0, 0, 1] directions. Solid lines are fit to the data with Gaussians as explained in the text. The dashed line indicates the background.}
\end{figure}

\section{Discussion}

In this section, we give an estimate of $I_{c}$ beyond the fit performed in section III that only gave the constraint 2$SI_{c}$ $<$ 20 meV and 2$S\Delta_{1}I_{c}$ $\approx$ 2500. Knowing that $I_{c} >> I_{1}, I_{2}$, it is tempting to describe the system as a set of weakly coupled ferromagnetic chains \cite{Steiner}. In such a model $I_{c} >> T_{Curie}$ and three dimensional order occurs for $T_{Curie} \approx 2S(S+1)\sqrt{I_{c}I_{1}}$ for 6 neighboring chains \cite{Scalapino}. However here our spin wave data show that 2S$I_{c}$ $<$ 20 meV. Knowing that 2$S$ $\sim$ 1 from Compton scattering, we deduce that $I_{c}$ is not much stronger than $T_{Curie}$ and the weakly coupled ferromagnetic chains model is therefore not valid here. For the purpose of an estimate, we therefore use the usual mean field approximation in its crude formulation, $T_{Curie} \approx I_{c}$, since $I_{c}$ is the dominant coupling (In such a case, the exact formula is $T_{Curie}=\frac{2z}{3}S(S+1)I_{c}$). This leads to the following order of magnitudes : $I_{c}$ $\approx$ 10 meV,  $\Delta_{1}$ $\approx$ 250 meV and $\Delta_{2}$ $\approx$ 0.02 meV. This is to be compared to the value of the out of plane anisotropy infered from the latest XAS results, $\Delta_{1}$ $\approx$ 50 meV \cite{Yamaguchi}. The origin of the X-Y nature of the system is the strong crystal field anisotropy. The dispersion relation given in formula (1) was used for exchange dominated planar ferromagnets like Tb \cite{Mackintosh} with in this case the use of total angular momentum $J$ instead of $S$. In such a case, the anisotropy gaps are expressed as            $\Delta_{1}$=6$JB^{0}_{2}$ and $\Delta_{2}$=36$J_{5}B^{6}_{6}$ with $J_{n}=J(J-1)...(J-n/2)$ and $B^{a}_{b}$ are the canonical coefficients of the crystal field Hamiltonian when expressed with the Stevens formalism \cite{Stevens}. We cannot use this formulation for CeRh$_{3}$B$_{2}$ since for Ce$^{3+}$ with a total momentum $J$=5/2 ground state, the coefficient $B^{6}_{6}$ is zero. As stated above, the gap at $q$=0 is a strong indication that in plane anisotropy exists. For Ce$^{3+}$ this is impossible with $J$=5/2. Hence the multiplet  $J$=7/2 has to be taken into account in the ground state wavefunction. This mixing between $J$=5/2 and $J$=5/2 in the ground state was first pointed out by the form factor measurements \cite{Givord1}. Concerning the anisotropy in the plane, one of the reported magnetization measurements at 2 K found a difference of 0.004 $\mu_{B}$ for the saturated magnetization between $a$ and $a^{*}$ axis \cite{Galatanu}. However this result was not reproduced by another group that also reports possible sample disorientation issue in the aforementioned work \cite{Givord2}. The small value of $\Delta_{2}$ that we proposed here implies that its signature will only occur at very low temperature probably much below 2 K.

\section{Conclusion}
In the present paper, we have focused on the magnetic interactions in CeRh$_{3}$B$_{2}$ while previous neutron scattering studies focused on single-site crystal-field contributions.
Usually rare-earth compounds are classified into exchange dominated or crystal field dominated systems \cite{Mackintosh}. This classification is clearly not valid for CeRh$_{3}$B$_{2}$ where all energy scales must be taken into account in the ground state and low-lying excitations: spin-orbit, crystal field and exchange. While our study does not answer the question of the microscopic origin of the strong Curie temperature, we clearly establish the huge anisotropy between the exchange along the $c$-axis and in the plane as well as the anisotropy within the plane that precludes a description of the magnetism in term of $J$=5/2 only. The spin-wave spectra extending presumably up to 80 meV at the $c$-axis zone boundary has the highest energy for any known cerium compound. This huge value is clearly a combination of crystal field anisotropy ($\Delta_{1}$) and exchange ($I_{c}$). The present work is aiming to stimulate further theoretical studies that would be able to treat on the same footing all the aspects (spin-orbit, crystal field and exchange) of the intriguing magnetism of CeRh$_{3}$B$_{2}$.

\section{Appendix}
In a neutron scattering experiment, the measured intensity is the convolution of the resolution function and the scattering function $S(\textbf{Q},\omega)$. This latter function is related to the imaginary part of the dynamical spin susceptibility $\chi"(\textbf{Q},\omega)$ via the fluctuation-dissipation theorem $S(\textbf{Q},\omega)=(1-exp(-\omega/T))^{-1}\chi"(\textbf{Q},\omega)$. In this paper, we use for the paramagnetic scattering (Fig.1a, Fig.2, Fig.5), a "quasielastic Lorentzian"  :
\begin{equation}
\chi"(\mathbf{q},\omega)=\frac{\chi_{q}\Gamma_{q}\omega}{\omega^{2}+\Gamma^{2}_{q}}
\end{equation}
For the spectra measured at $\textbf{Q}$=(0, 0, 1) (Fig.1b), we use the following form, that was named "inelastic Lorentzian" in the paper :
\begin{equation}
\chi"(\mathbf{q},\omega)=\frac{1}{2}\left[\frac{\chi_{q}\Gamma_{q}\omega}{(\omega-\omega_{0})^{2}+\Gamma^{2}_{q}}+\frac{\chi_{q}\Gamma_{q}\omega}{(\omega+\omega_{0})^{2}+\Gamma^{2}_{q}}\right]
\end{equation}
This form is phenomenologically used to reproduce the broadening of the peak. The origin of this anomalous shape is presumably the steep dispersion along the $c$-axis that is picked up by the finite resolution function. This effect is strongest at the minimum of the dispersion. For the spin wave spectra at finite $\textbf{q}$, we use a gaussian for $S(\textbf{Q},\omega)$ both for the constant $\textbf{Q}$ (Fig.2) and constant $\omega$ (Fig.3) scans, i.e for $x$=h,l or $\omega$.
\begin{equation}
S(\mathbf{q},\omega)=S(x)=I.exp(-ln2((x-C)/W/2)^2)
\end{equation}
with $I$ the peak intensity, $C$ the position of the peak and $W$ its Full Width at Half Maximum (in $\textbf{Q}$ or $\omega$ space).

\end{document}